\documentclass[aip,reprint]{revtex4-1}
\usepackage{amsmath,amsfonts}

\begin{document}


\title{Analytic properties of the electromagnetic 
Green's function} 



\author{Boris Gralak}
\email[]{boris.gralak@fresnel.fr}
\affiliation{CNRS, Aix-Marseille Universit{\'e}, Centrale Marseille, 
Institut Fresnel, UMR 7249, 13013 Marseille, France}


\date{\today}

\begin{abstract}
The electromagnetic Green's function is expressed from the inverse 
Helmholtz operator, where a second frequency has been introduced as a new degree of freedom. 
The first frequency results from the frequency decomposition of the electromagnetic field 
while the second frequency is associated with the dispersion of the dielectric permittivity. 
Then, it is shown that the electromagnetic Green's function is analytic with respect to 
these two complex frequencies as soon as they have positive imaginary part. Such analytic 
properties are also extended to complex wavevectors. Next, Kramers-Kronig expressions for 
the inverse Helmholtz operator and the electromagnetic Green's function are derived. In 
addition, these Kramers-Kronig expressions are shown to correspond to the classical 
eigengenmodes expansion of the Green's function established in simple situations. Finally, 
the second frequency introduced as a new degree of freedom is exploited to characterize
non-dispersive systems.
\end{abstract}

\pacs{}

\maketitle 

%
%
\def\x{\text{\bfseries\sffamily\textit{x}}}
\def\xp{\text{\bfseries\sffamily\textit{x}}_{\!\text{\sffamily\textit{p}}}}
\def\xs{\text{\sffamily\textit{x}}}
\def\y{\text{\bfseries\sffamily\textit{y}}}
\def\k{\text{\bfseries\sffamily\textit{k}}}
\def\kp{\text{\bfseries\sffamily\textit{k}}_{\!\text{\sffamily\textit{p}}}}
\def\u{\text{\bfseries\sffamily\textit{u}}}

\def\E{\text{\bfseries\sffamily\textit{E}}}
\def\P{\text{\bfseries\sffamily\textit{P}}}
\def\B{\text{\bfseries\sffamily\textit{B}}}
\def\D{\text{\bfseries\sffamily\textit{D}}}
\def\A{\text{\bfseries\sffamily\textit{A}}}
\def\J{\text{\bfseries\sffamily\textit{J}}}
\def\F{\text{\sffamily\textit{F}}}
\def\t{\text{\sffamily\textit{t}}}
\def\t{t}
\def\s{\text{\sffamily\textit{s}}}
\def\nui{\text{\sffamily\textit{$\nu$}}}
\def\f{\text{\sffamily{f}}}
\def\c{\text{\sffamily\textit{c}}}
\def\v{\text{\sffamily\textit{v}}}
\def\ks{\text{\sffamily\textit{k}}}
\def\rot{\boldsymbol{\nabla \times }}
\def\dt{\partial_{\t}}
\def\hatLE{\hspace*{0.35mm}\widehat{\hspace*{-0.35mm}\text{\bfseries\sffamily\textit{E}}\hspace*{0.35mm}}\hspace*{-0.35mm}}
\def\LJ{\,\widehat{\!\text{\bfseries\sffamily\textit{J}}\,}\!}
\def\LD{\,\widehat{\!\text{\bfseries\sffamily\textit{D}}\,}\!}
\def\E{\text{\bfseries\textit{E}}}
\def\P{\text{\bfseries\textit{P}}}
\def\B{\text{\bfseries\textit{B}}}
\def\H{\text{\bfseries\textit{H}}}
\def\D{\text{\bfseries\textit{D}}}
\def\A{\text{\bfseries\textit{A}}}
\def\J{\text{\bfseries\textit{J}}}
\def\F{\text{\textit{F}}}
\def\LE{\text{\bfseries\sffamily\textit{{E}}}}
\def\LP{\text{\bfseries\sffamily\textit{P}}}
\def\LB{\text{\bfseries\sffamily\textit{B}}}
\def\LH{\text{\bfseries\sffamily\textit{H}}}
\def\LD{\text{\bfseries\sffamily\textit{D}}}
\def\LA{\text{\bfseries\sffamily\textit{A}}}
\def\LJ{\text{\bfseries\sffamily\textit{J}}}
\def\LF{\text{\sffamily\textit{F}}}
\def\LS{\text{\sffamily\textit{S}}}
\def\vS{\text{\bfseries\sffamily\textit{S}}}
\def\ep{\varepsilon}
\def\om{\omega}
\def\x{\text{\bfseries\sffamily\textit{x}}}
\def\y{\text{\bfseries\sffamily\textit{y}}}
\def\k{\text{\bfseries\sffamily\textit{k}}}

\def\rot{\partial_{\x} \!\times\!}
\def\dint{\displaystyle\int}
\def\R{\mathbb{R}}
\def\Proj{\mathsf{P}}
\def\He{\mathsf{H}}
\def\Res{\mathsf{R}}
\def\Re{\mathsf{R}_e}
\def\G{\mathsf{G}}
\section{Introduction\label{sec1}}
The electromagnetic Green's function is a fundamental quantity in the analysis 
of systems described by macroscopic Maxwell's equations \cite{Jackson}. It is 
defined from the inverse of the Helmholtz operator \cite{RS3}, which provides the 
electric field radiated by a current source. For a medium described by the 
electric permittivity $\ep(\x,z)$ depending on the space variable $\x \in \R^3$ 
and the complex frequency $z$, the Helmholtz operator $\He_e(z)$ is given by
\begin{equation}
\big[ \He_e(z) \LE \big] (\x) = z^2 \ep(\x,z) \mu_0 \LE(\x) - \rot \rot \LE(\x) \, ,
\label{Hez}
\end{equation}
where $\partial_{\x} \times$ is the curl operator, and $\mu_0$ the vacuum permeability. 
Then, the electromagnetic Green's function can be defined from the inverse of the 
Helmholtz operator by
\begin{equation}
\big[ \He_e(z)^{-1} \vS \big] (\x) = 
\dint_{\R^3} d\y \, \G_e(\x,\y;z) \cdot \vS(\y) \, , 
\label{Green-def}
\end{equation}
where $\vS(\x)$ is proportional to a current source density. In this paper, 
analytic properties of the electromagnetic Green's function $\G_e(\x,\y;z)$ 
are rigorously deduced from those of the inverse Helmholtz operator (\ref{Hez}). 

It is well-known that the Green's function $\G_e(\x,\y;z)$ is an analytic function in 
the upper half space of complex frequencies $z$. This is a direct consequence of the causality 
principle and passivity \cite{Jackson,Landau8}. Notice that the frequency dependence of the 
electromagnetic Green's function has two different origins in Maxwell's equations: the first 
one is the consequence of the frequency decomposition of the time derivative of the fields
in Maxwell's equations, and the second one is the frequency dispersion which results in 
the frequency dependence of the permittivity $\ep(\x,z)$. Here, it is proposed to show that 
the analytic properties of $\G_e(\x,\y;z)$ can be established for these two frequencies 
independently (section \ref{sec3}). In particular, this is exploited in section \ref{sec5} to 
provide a rigorous proof of the analyticity and causality in the non-dispersive case. 
Also, in the cases of homogeneous or periodic geometry, the analytic properties are 
extended to complex wavevectors (section \ref{sec6}). 

The analytic properties of the electromagnetic Green's function can be used to compute
Sommerfeld integrals and time-dependent electromagnetic fields \cite{GM12}, for instance
defining analytic continuation in the plane of complex frequencies. Also, the Kramers-Kronig 
relations are based on such properties of analyticity \cite{Jackson,Landau8}. For instance, 
new Kramers-Kronig relations have been established in reference \cite{Gra15} for the 
reflection and transmission coefficients in non-normal incidence. In this paper, it is 
proposed in section \ref{sec4} to derive Kramers-Kronig expressions for the inverse Helmholtz 
operator and the electromagnetic Green's function. Arguments are provided to interpret these 
expressions as generalizations to dispersive and absorptive systems of the well-known 
eigenmodes expansion \cite{RemiC15} established for simple closed cavity (without dispersion 
and absorption).
\section{Generalized Helmholtz operator\label{sec2}}
\subsection{Maxwell's equations}
We start with Maxwell's equations in dielectric media. Let $\E(\x,\t)$, $\H(\x,\t)$ 
and $\P(\x,\t)$ be respectively  the time-dependent electric, magnetic and 
polarization fields. Then, equations of macroscopic electromagnetics \cite{Jackson} are
\begin{equation}
\begin{array}{l}
\ep_0 \, \dt \E(\x,\t) + \dt \P(\x,\t) = \rot \H(\x,\t)  
- \J(\x,\t) \, , \\[2mm]
\mu_0 \, \dt \H(\x,\t) = - \rot\E(\x,\t) \, , 
\end{array}
\label{Max-eq}
\end{equation}
where $\dt$ is the partial derivative with respect to time, 
and $\J(\x,\t)$ is the current source density. 
In addition, the electric field is related to the polarization 
through the constitutive equation
\begin{equation}
\P(\x,\t) = \displaystyle\int_{-\infty}^{\t} d \s \, \chi(\x,\t-\s) \E(\x,\s) \, , 
\label{const-eq}
\end{equation}
where $\chi(\x,\t)$ is the electric susceptibility. A Fourier decomposition with 
respect to the time of the equations above leads to 
\begin{equation}
\begin{array}{l}
- i \om \ep(\x,\om) \, \LE(\x,\om) = \rot \LH(\x,\om) - \LJ(\x,\om) \, , \\[2mm]
- i \om \mu_0 \LH(\x,\om) = - \rot\LE(\x,\om) \, ,
\end{array}
\label{Max-om}
\end{equation}
where the dielectric permittivity is defined as
\begin{equation}
\ep(\x,\om) - \ep_0 = \displaystyle\int_0^{\infty} d \t \, 
\exp[i \om \t] \, \chi(\x,\t) \, . 
\label{def-ep}
\end{equation}
Here, according to the causality principle, it has been used that the susceptibility 
$\chi(\x,\t)$ vanishes for negative times, i.e. $\chi(\x,\t) = 0$ if $\t<0$. 
Also, without loss of generality, it is assumed that the susceptibility 
$\chi(\x,\t)$ must be bounded with respect to the time. 
Consequently, $t$ is always positive in the integral above, and the permittivity 
remains well-defined if the 
real frequency is replaced by the complex frequency $z = \om + i \eta$ with positive 
imaginary part Im$(z) = \eta >0$. Moreover, its derivative with respect to the complex 
frequency is well-defined since the function in the integral 
\begin{equation}
\dfrac{\partial \ep}{\partial z}(\x, z) = 
\displaystyle\int_0^{\infty} d \t \, (i\t) \exp[i z \t] \chi(\x,\t) \,  
\label{analytic-ep}
\end{equation}
has exponential decay for Im$(z)>0$. It follows that the permittivity $\ep(\x,z)$ 
is an analytic function in the half plane of complex frequencies $z$ with positive 
imaginary part, which will be denominated by ``upper half plane'' from now on. 
Finally, the susceptibility can be retrieved through the inverse Fourier 
(Laplace) transform
\begin{equation}
\chi(\x,\t) = \dfrac{1}{2\pi} \displaystyle\int_{\Gamma_\eta} dz \, \exp[-i z \t] \, 
[ \ep(\x,z) - \ep_0 ] \, ,
\label{chi-ep}
\end{equation}
where $\Gamma_\eta$ is an horizontal line of complex numbers $z = \om + i \eta$ 
with $\om \in \R$, at a distance $\eta$ from the real axis.

The Helmholtz equation is directly deduced from the set of 
equations (\ref{Max-om}), where the $\om$-dependence of the fields 
has been omitted:
\begin{equation}
\begin{array}{ll}
\big[ \He_e(\om) \LE \big] (\x) & \hspace*{-1mm}= 
\om^2 \ep(\x,\om) \mu_0 \LE(\x) - \rot \rot \LE(\x) \\[2mm]
& \hspace*{-1mm}= - i \om \mu_0 \LJ(\x) \, .
\end{array}
\label{Heom}
\end{equation}
Since the permittivity is well-defined in the upper half plane Im$(z)>0$, 
the definition of the Helmholtz operator $\He_e(\om)$ can be 
also extended to all complex frequency $z$ with Im$(z)>0$. It can be shown 
rigorously that the inverse of $\He_e(z)$ exists and is analytic in this domain 
Im$z>0$ using the the auxiliary field formalism \cite{Tip98}. Indeed, 
adding a new ``auxiliary'' field $\A(\t)$ to the electromagnetic field 
to form the total vector field $F(\t) = [ \E(\t), \H(\t), \A(t)]$, the set of 
Maxwell's equations (\ref{Max-eq}) [without the source $\J(\x,t)$] can be written 
as the unitary time-evolution equation $\dt \F(\t) = - i \mathsf{K} \F(\t)$, 
where $\mathsf{K}$ is a time-independent selfadjoint operator. 
%
The inverse $[z - \mathsf{K}]^{-1}$ is then well-defined for all complex number 
$z$ with Im$(z)>0$, and is moreover an analytic function of $z$. Next, the 
inverse of the Helmholtz operator is retrieved by projecting the total field 
$\LF(z)$ on the electric fields $\LE(z)$, and using the Feshbach projection 
formula \cite{Tip00,GT10}. Since the projector on electric fields is 
$z$-independent, the inverse of the Helmholtz operator has the same analytic 
properties as the inverse $[z - \mathsf{K}]^{-1}$. 

In this paper, it is proposed to use arguments based on the properties 
of the permittivity $\ep(\x,z)$, and then to transpose them directly to the 
Helmholtz operator and the electromagnetic Green's function.

\subsection{The permittivity}
Properties of the permittivity are derived in this subsection. The key point
is the generalized expression of Kramers-Kronig relations 
\cite{Tip98,LGG13} for the permittivity:
\begin{equation}
\ep(\x,z) = \ep_0 - \dint_{\R} d\nu \dfrac{\sigma(\x,\nu)}{z^2 - \nu^2} \, ,
\label{KKR}
\end{equation}
where
\begin{equation}
\sigma(\x,\nu) = \text{Im} \, \displaystyle\frac{\nu [ \ep(\x,\nu) - \ep_0] }{\pi} 
\ge 0 \, .
\label{sigma}
\end{equation}
One can checked that, when the limit $\text{Im}(z) = \eta \downarrow 0$ is 
considered in (\ref{KKR}), a Dirac function appears in the integral (\ref{KKR})
and the relation (\ref{sigma}) is retrieved.
Notice that it has been assumed that only passive media are considered. Under this 
condition, the electromagnetic energy must decrease with time, and thus the
permittivity must have positive imaginary part \cite{Landau8}: hence the
function $\sigma(\x,\nu)$ is positive. At the microscopic scale, this function 
corresponds to the oscillator strength \cite{CDL2} which must be positive. 
Notice that this passivity requirement(\ref{sigma}) can be extended to all 
complex frequency in the upper half plane. Indeed, using that 
$\sigma(\x,\nu) = \sigma(\x,-\nu)$, the expression (\ref{KKR}) 
can be written as
\begin{equation}
z[ \ep(\x,z) - \ep_0 ] = - 
\dint_{\R} d\nu \, \dfrac{\sigma(\x,\nu)}{z - \nu} \, , 
\label{KKR2}
\end{equation}
which implies
\begin{equation}
\text{Im} \{ z [ \ep(\x,z) - \ep_0 ] \} 
= \text{Im}(z) \dint_{\R} d\nu \, \dfrac{\sigma(\x,\nu)}{| z - \nu |^2} \, 
\ge 0 \, . 
\label{passivity2}
\end{equation}
The derivative $( \dt \chi ) (\x,t)$ of the susceptibility corresponds to the 
microscopic currents and must be bounded since the charges have their
velocity bellow the one of light. It can be calculated using the 
Kramers-Kronig relation 
(\ref{KKR}) or (\ref{KKR2}):
\begin{equation}
\begin{array}{ll}
( \dt \chi ) (\x,t) \hspace*{-1mm}& = \dfrac{1}{2 \pi} \dint_{\Gamma_\eta}
dz \, \exp[-i z t] \, (-iz) \, [ \ep(\x,z) - \ep_0 ] \\[4mm]
& = \dfrac{1}{2 \pi} \dint_{\R} d\nu \, \sigma(\x,\nu) 
\dint_{\Gamma} dz \, (i) \, \dfrac{\exp[-i z t]}{z - \nu} \, .
\end{array}
\label{finite}
\end{equation}
The integral above is computed by closing the line $\Gamma$ by a semi circle 
in the upper (for $\t<0$) or lower (for $\t>0$) half spaces. It is retrieved 
that the susceptibility $\chi(\x,\t)$ vanishes for $\t<0$ and, for $\t>0$:
\begin{equation}
\begin{array}{rl}
( \dt \chi ) (\x,t) & = \dint_\R d\nu \, \sigma(\x,\nu) \cos[\nu\t] \, .
\end{array}
\end{equation}
This implies in particular that the integral of the function $\sigma(\x,\nu)$ 
is finite since
\begin{equation}
\dint_{\R} d\nu \, \sigma(\x,\nu) = [ \dt \chi ] (\x,0^+) < \infty \, . 
\label{finite}
\end{equation}
Also, it can be checked that $[ \dt \chi ] (\x,t)$ is continuous of $\t$ 
in the general case (except at $\t = 0$), 
and that $[ \dt \chi ] (\x,\t)$ is bounded by $[\dt \chi](\x, 0)$. 

Finally, an important estimate of the permittivity is provided at the limit 
of high frequencies. The function $[ \dt \chi ] (\x,t)$ being bounded and 
continuous (except at $t=0$), the second derivative $[ \dt^2 \chi ] (\x,t)$
can be defined for all $t \neq 0$. Hence the following relation holds
\begin{equation}
z^2[ \ep(\x,z) - \ep_0 ] = - [\dt \chi](\x,0^+) 
- \dint_0^\infty \hspace*{-3mm}d\t \, \exp[iz\t] \, [\dt^2 \chi] (\x,\t) \, .    
\label{z2ep}
\end{equation}
Since the derivative $[ \dt \chi ] (\x,t)$ corresponds to the microscopic 
currents, it is related to the impulsion of the charges: it is then reasonable to 
assume that its variations are bounded because of the inertia (charges have a mass), 
otherwise an infinite power is required. Notice that this argument does not apply to 
the initial time $\t = 0$, where the causality requirement switches on (or switches off) 
instantaneously a force on the charges. 
Thus, from now on, it is assumed that the second derivative $[\dt^2 \chi] (\x,\t) $ is 
bounded for $t > 0$. Using that the integral in the relation (\ref{z2ep}) 
is the Fourier transform of an integrable function, this relation 
implies for fixed $\eta$ in $z = \om + i \eta$ the important estimate
\begin{equation}
z^2[ \ep(\x,z) - \ep_0 ] 
\underset{\om \rightarrow \infty}{\longrightarrow} - [\dt \chi](\x,0^+) \, .
\label{z2epinfty}
\end{equation}


\subsection{Helmholtz operator and extension of its definition\label{sec2C}}
In order to define rigorously the inverse of the Helmholtz operator for 
complex frequencies $z$, the set of equation (\ref{Max-om}) is written as
\begin{equation}
\big[ \mathsf{M}_0(z) + \mathsf{V}(\x,z) \big] \LF(\x) 
= \LS(\x) \, ,
\label{Max-z}
\end{equation}
where $\LF(\x)$ contains the electromagnetic field and $\LS(\x)$ the current source: 
\begin{equation}
\LF(\x) = \left[ \begin{array}{c}
\LE(\x) \\[1mm] \LH(\x) \end{array} \right] \, , \quad 
\LS(\x) = \left[ \begin{array}{c}
- i \ep_0^{-1} \LJ(\x) \\[1mm] 0 \end{array} \right] \, .
\label{Max-z-def}
\end{equation}
The operator $\mathsf{M}_0(z)$ corresponds to Maxwell's equations in vacuum and 
$\mathsf{V}(\x,z)$ contains the response of the material:
\begin{equation}
\begin{array}{rl}
\mathsf{M}_0(z) & \hspace*{-1mm}= \left[ \begin{array}{lr}
z & i\ep_0^{-1}\partial_{\x} \times \\[1mm] - i \mu_0^{-1}\partial_{\x} \times & z  \end{array} \right] \, , \\[6mm]
\mathsf{V}(\x,z) & \hspace*{-1mm}= \left[ \begin{array}{lr}
z \{\ep(\x,z)/ \ep_0 - 1\} & 0 \\[1mm] 0 & 0 \end{array} \right] \, .
\end{array}
\label{Max-def}
\end{equation}
Let $\langle \cdot , \cdot \rangle$ be the standard inner product in the Hilbert space 
of square integrable electromagnetic fields:
\begin{equation}
\big\langle \LF_1 , \LF_2 \big\rangle = 
\dint_{\R^3} d\x \big[ \ep_0 \overline{\LE_1(\x)} \cdot \LE_2(\x) + \mu_0
\overline{\LH_1(\x)} \cdot \LH_2(\x) \big] \, .
\end{equation}
If for all field $\LF$ there exists a constant $\alpha> 0 $ such that 
\begin{equation}
\big| \big\langle \LF , \big[ \mathsf{M}_0(z) + \mathsf{V}(\x,z) \big] 
\LF \big\rangle \big| \ge \alpha \, \big\langle \LF , \LF \big\rangle \, ,
\end{equation}
then the operator $[ \mathsf{M}_0(z) + \mathsf{V}(\x,z) ]$ is invertible, 
and its inverse is bounded by $\alpha^{-1}$. Using that the curl is selfadjoint, 
the following relationship is obtained
\begin{equation}
\begin{array}{ll}
\text{Im} \, \big\langle \LF , \mathsf{M}_0(z) \LF \big\rangle & \hspace*{-1mm}= 
\text{Im}(z) \dint_{\R^3} d\x \big[ \ep_0 |\LE(\x)|^2 + 
\mu_0 |\LH(\x)|^2 \big] \\[4mm]
& \hspace*{-1mm}= \text{Im}(z) \, \big\langle \LF , \LF \big\rangle \, .
\end{array}
\end{equation}
The term with $\mathsf{V}(\x,z)$ containing the relative permittivity is 
estimated using the extended passivity requirement (\ref{passivity2}):
\begin{equation}
\text{Im} \, \big\langle \LF , \mathsf{V}(\x,z) \LF \big\rangle = \!
\dint_{\R^3} \hspace*{-2mm} d\x \, |\LE(\x)|^2 
\text{Im} \{ z[ \ep(\x,z) - \ep_0 ] \} \ge 0 \, . 
\label{ImV}
\end{equation}
The combination of the two equations leads to
\begin{equation}
\big| \big\langle \LF , \big[ \mathsf{M}_0(z) + \mathsf{V}(\x,z) \big] \LF \big\rangle \big| \ge 
\text{Im}(z) \,  \big\langle \LF , \LF \big\rangle \, ,
\label{bound}
\end{equation}
which implies that the inverse $[ \mathsf{M}_0(z) + \mathsf{V}(\x,z) ]^{-1}$ 
is well-defined for Im$(z)>0$ and bounded by 
$\alpha^{-1} = 1/ \text{Im}(z) $. In addition, this inverse is an analytic 
function of the complex frequency $z$ for Im$(z)>0$ as well as the 
permittivity function $\ep(\x,z)$ in 
$\mathsf{V}(\x,z)$, and $\mathsf{M}_0(z)$. The analyticity property can be also shown using the 
first resolvent formula \cite{RS1}. Denoting $\mathsf{M}(z) = 
\mathsf{M}_0(z)+\mathsf{V}(\x,z)$, the difference of the inverses at 
$z$ and $z_0$ is
\begin{equation}
\mathsf{M}(z)^{-1} - \mathsf{M}(z_0)^{-1} = 
\mathsf{M}(z)^{-1} (z_0-z) \mathsf{A} \, \mathsf{M}(z_0)^{-1} \, ,
\label{ana1}
\end{equation}
where, using expression (\ref{KKR2}) for $\mathsf{V}(\x,z)$, 
\begin{equation}
\mathsf{A} = \left[ \begin{array}{lr}
1 + m & 0 \\ 0& 1 \end{array} \right] \, , 
\quad m = \ep_0^{-1}\dint_{\R} d\nu \dfrac{\sigma(\x,\nu)}{(z-\nu)(z_0-\nu)} .
\label{ana2}
\end{equation}
The operator $\mathsf{A}$ is bounded thanks to the condition (\ref{finite}), and 
$\mathsf{M}(z_0)^{-1}$ is bounded by $1 / \text{Im}(z_0) $.
The identity (\ref{ana1}) implies
\begin{equation}
\begin{array}{ll}
\mathsf{M}(z)^{-1} & \hspace*{-1mm} = \mathsf{M}(z_0)^{-1} \big[ 1 - 
(z_0-z) \mathsf{A} \mathsf{M}(z_0)^{-1} \big]^{-1} \\[2mm]
& \hspace*{-1mm} = \mathsf{M}(z_0)^{-1} \Big\{ 1+\displaystyle\sum_{p}
(z_0-z)^p \big[\mathsf{A} \mathsf{M}(z_0)^{-1} \big]^{p} \Big\} \, .
\end{array}
\label{ana3}
\end{equation}
The last series converges in norm provided $|z-z_0|$ is smaller than the 
inverse of the nom of $[\mathsf{A} \mathsf{M}(z_0)^{-1} ]$. This power series expansion
shows that the inverse $[ \mathsf{M}_0(z) + \mathsf{V}(\x,z) ]^{-1}$ is an
analytic function if Im$(z_0)>0$.

In the last step, the Helmholtz operator is retrieved using the projector 
on electric fields $\Proj$, defined by $\Proj \LF(\x) = \LE(\x)$. Then, 
equation (\ref{Max-z}) yields 
\begin{equation}
\begin{array}{ll}
\LE(\x) &\hspace*{-1mm} = \Proj \LF(\x) = \Proj \big[ \mathsf{M}_0(z) + 
\mathsf{V}(\x,z) \big]^{-1} \LS(\x) \\[2mm]
&\hspace*{-1mm} = - i \ep_0^{-1} \Proj \big[ \mathsf{M}_0(z) + 
\mathsf{V}(\x,z) \big]^{-1} \Proj \LS(\x) \, , 
\end{array}
\end{equation}
where, according to (\ref{Max-z-def}), $\LS(\x) = \Proj \LS(\x) $ 
has the single ``electric component'' 
$-i \ep_0^{-1} \LJ(\x)$. 
According to (\ref{Heom}), the inverse Helmholtz operator is given by 
$\LE(\x) = - i z \mu_0 \He_e(z)^{-1} \LJ(\x)$, and the comparison with 
the equation above provides 
\begin{equation}
\He_e(z)^{-1} = \dfrac{1}{z \mu_0 \ep_0} 
\Proj \big[ \mathsf{M}_0(z) + \mathsf{V}(\x,z) \big]^{-1} \Proj \, , 
\label{invHelm}
\end{equation}
This expression shows that all the properties of the inverse 
$[ \mathsf{M}_0(z) + \mathsf{V}(\x,z) ]^{-1}$ are directly 
transposable to the inverse Helmholtz operator. In particular 
the inverse Helmholtz operator is bounded by
\begin{equation}
\big\Arrowvert \He_e(z)^{-1} \big\Arrowvert \leq \dfrac{1}{|z|\ep_0\mu_0 \text{Im}(z)} \, .
\label{boundinvHe}
\end{equation}

It is stressed that the bound $\alpha = \text{Im}(z)$ in (\ref{bound}) 
is governed by the imaginary part of $z$ in $\mathsf{M}_0(z)$ only, and 
thus is independent of the complex number $z$ in $\mathsf{V}(\x,z)$. Hence the 
possibility to consider the two complex frequencies in $\mathsf{M}_0(z)$ and 
$\mathsf{V}(\x,z)$ independently. As a result, it is obtained that the inverse 
\begin{equation}
\big[ \mathsf{M}_0(z) + \mathsf{V}(\x,\xi) \big]^{-1} \le 
[\, \text{Im}(z) \,]^{-1} \, , 
\label{Rezxi}
\end{equation}
exists and is analytic with respect to both complex frequencies $z$ and $\xi$ 
as soon as Im$(z)>0$ and Im$(\xi)>0$. In particular power series expansions like 
(\ref{ana3}) can be stated for both variables $z$ and $\xi$ independently. 
This property can be transposed to the inverse of a 
generalized version of the Helmholtz operator. The expression of this operator, denoted 
by $\He(z,\xi)$, can be obtained by replacing $\mathsf{V}(\x,z)$ by 
$\mathsf{V}(\x,\xi)$ and then by eliminating the magnetic field $\LH(\x)$ in equation 
(\ref{Max-z}):
\begin{equation}
\He(z,\xi) = z^2 \ep_0 \mu_0 + z \mu_0 \xi [\ep(\x,\xi) - \ep_0] - \rot \rot
\label{Helmzxi}
\end{equation}
A relation similar to (\ref{invHelm}) shows that the inverse of $\He(z,\xi)$ 
exists and is analytic of both complex variables $z$ and $\xi$ with 
Im$(z)>0$ and Im$(\xi)>0$. It is bounded by 
\begin{equation}
\big\Arrowvert \He(z,\xi)^{-1} \big\Arrowvert \leq \dfrac{1}{|z|\ep_0\mu_0 \text{Im}(z)} \, .
\label{Helmzxi}
\end{equation}
This extended definition of Helmholtz operator is used in section \ref{sec5} 
to analyze non dispersive systems. 

\section{The electromagnetic Green's function\label{sec3}}

The electromagnetic Green's function can be introduced from the 
inverse Helmholtz operator as shown by equation (\ref{Green-def}). 
While the left side of the equation, $\He_e(z)^{-1} \vS$, is 
well-defined, there is no argument which ensures the existence of the 
Green's function in the right side in Electromagnetism. Indeed, the 
existence of the Green's function is usually the consequence of 
the compact or Hilbert-Schmidt nature of the corresponding operator \cite{RS1}. 
This compact nature could have been obtained by considering the difference 
of the original inverse $[ \mathsf{M}_0(z) + \mathsf{V}(\x,z)]^{-1}$ 
with the free inverse $\mathsf{M}_0(z)^{-1}$ [see reference \cite{RS4}]. However, 
in the case of Electromagnetism, this technique is not suitable because of the 
presence of the ``\textit{static}'' modes which generate a ``Dirac'' 
singularity in the Green's function (see reference \cite{GGT07} for 
investigations on the singularity). 
In practice, for square integrable functions $\phi$ and $\psi$, it is always 
possible to define coefficients like
\begin{equation}
\big\langle \phi, \He_e(z)^{-1} \psi \big\rangle 
=\dint_{\R^3} d\x \, \overline{\phi(\x)} \, \big[ \He_e(z)^{-1} \psi \big] (\x)
\label{Res-weak}
\end{equation}
corresponding to
\begin{equation}
\dint_{\R^3} d\x \dint_{\R^3} d\y \,\, \overline{\phi(\x)} 
\G_e(\x,\y;z) \psi(\y) \, .
\label{Green-weak}
\end{equation}
[From now, the brakets $\langle \cdot , \cdot \rangle$ denotes the standard 
inner product (\ref{Res-weak}) for the solely electric fields.]
The functions $\phi$ and $\psi$ can be the elements of an orthonormal
basis $\{ \phi_p(\x) \}$ of the square integrable functions. Then, the 
coefficients $\big\langle \phi_p, \He_e(z)^{-1} \phi_q \big\rangle$ 
can be used to define ``formally'' 
\begin{equation}
\G_e(\x,\y;z) = \displaystyle\sum_{n,m} 
\big\langle \phi_p, \He_e(z)^{-1} \phi_q \big\rangle
\phi_p(\x) \otimes \overline{\phi_q(\y)} \, ,
\label{Green-basis}
\end{equation}
where the symbol $\otimes$ means that the tensor product is considered. 
Note that this definition is only ``formal'' because there is no argument ensuring 
that the convergence of the sum in the case $\He_e(z)^{-1}$ is not Hilbert-Schmidt. 
The functions $\phi$ and $\psi$ in (\ref{Res-weak}) and (\ref{Green-weak}) can be 
also chosen to approach the identity. Let he function $\phi_a$ be defined by 
$\phi_{a}(\x) = (3/4 \pi) a^{-3}$ if $|\x| \le a$  
and $\phi_{a}(\x) = 0$ if $|\x| \ge a$ in order to approach the Dirac function 
$\delta(\x)$ when $a \downarrow 0$. Then, for $a$ small enough, 
$\phi_{\x_0}(\x) = \phi_a(\x-\x_0) \approx \delta(\x - \x_0)$
and $\phi_{\y_0}(\y) = \phi_a(\y-\y_0) \approx \delta(\y - \y_0)$, and the coefficient 
$\big\langle \phi_{\x_0}, \He_e(z)^{-1} \phi_{\y_0} \big\rangle$ can approach the Green's 
function:
\begin{equation}
\begin{array}{l}
\big\langle \phi_{\x_0}, \He_e(z)^{-1} \phi_{\y_0} \big\rangle \\[2mm]
\quad \quad = \dint_{\R^3} \hspace*{-1mm}d\x \dint_{\R^3} \hspace*{-1mm}d\y \, 
\phi_a(\x-\x_0) \G_e(\x,\y;z) \phi_a(\y-\y_0) \\[4mm]
\quad \quad \approx \G_e(\x_0,\y_0;z) \, .
\end{array}
\label{Green-local}
\end{equation}
Here, it is stressed that nothing authorizes to take the limit $a \downarrow 0$ 
since the function $\phi_a$ is not square integrable at this limit. Thus the 
coefficient $\big\langle \phi_{\x_0}, \He_e(z)^{-1} \phi_{\y_0} \big\rangle$ 
just allows to address an approximation of the Green's function $\G_e(\x_0,\y_0;z)$. 

According to these arguments above, it is assumed that the electromagnetic 
Green's function can be defined, and that its properties can be established from 
the coefficients $\langle \phi, \He_e(z)^{-1} \psi \rangle$. 
First, it is clear that all the analytic properties of $\He_e(z)^{-1}$ 
and $\He(z,\xi)^{-1}$ are directly transposable to coefficients 
$\langle \phi, \He_e(z)^{-1} \psi \rangle$ and 
$\langle \phi, \He(z,\xi)^{-1} \psi \rangle$. Indeed, it is enough to 
expand the inverse in $\langle \phi, \He_e(z)^{-1} \psi \rangle$ 
in a power series like (\ref{ana3}) and then to check that the resulting power 
series with the coefficients converges. Another important properties of the 
Green's function is the behavior for large frequency $z$. First, it is 
shown in the appendix \ref{A1} that it decreases like $1/(z^2 \ep_0 \mu_0)$ or, 
equivalently, 
\begin{equation}
\lim_{|z| \rightarrow \infty} \, z^2 \ep_0 \mu_0 \, \big\langle \phi, 
\He_e(z)^{-1} \psi \big\rangle =  \big\langle \phi, \psi \big\rangle \, .
\label{limzinfty}
\end{equation}
It has to be noticed that this asymptotic behavior is for the modulus $|z|$ of the complex 
frequency which tends to infinity. An important asymptotic regime is the limit $\om \rightarrow \infty$
in the complex frequency $z$, i.e. for fixed imaginary part Im$(z)$.
In this case, one can show that 
\begin{equation}
z^2\big[ \He_e(z)^{-1} - \He_0(z)^{-1} \big] 
\label{idimp}
\end{equation}
is bounded when $\om \rightarrow \infty$. This can be established writting 
the difference
\begin{equation}
\He_e(z)^{-1} - \He_0(z)^{-1} = - \He_0(z)^{-1} z^2\mu_0 [\ep(\x,z) - \ep_0] 
\He_e(z)^{-1} \, ,
\label{id2}
\end{equation}
and then using the bound (\ref{boundinvHe}) and the estimate (\ref{z2epinfty}). 
Hence it is found that 
\begin{equation}
\begin{array}{lcl}
z^2\big[ \He_e(z)^{-1} \hspace{-3mm} & - & \hspace{-3mm} \He_0(z)^{-1} \big] \\[3mm] 
& \underset{ \om \rightarrow \infty}{\approx} &
z \He_0(z)^{-1} \, [\dt \chi](\x,0^+) \,  z \He_e(z)^{-1} \\[2mm]
& \leq & \dfrac{[\dt \chi](\x,0^+)}{[\ep_0 \mu_0 \, \text{Im}(z)]^2} \,  .
 \end{array}
\label{Hzinfty}
\end{equation}

These properties and asymptotic behaviors will be used in the next section to derive 
a version of Kramers-Kronig relations for the electromagnetic Green's function. 

\section{Kramers-Kronig relations for the Electromagnetic Green's function\label{sec4}}

The Kramers-Kronig relations can be applied to all function derived from a 
causal signal. It is generally used to analyze the dielectric permittivity, 
the permeability or the optical index \cite{Jackson,Landau8}. A new version 
of Kramers-Kronig relations, given by equation (\ref{KKR}), has been proposed 
recently \cite{Tip98,LGG13}. This version shows that the general expression 
of the permittivity is a continuous superposition of elementary resonances 
given by the elastically bound electron model. Thus it extends 
the classical Drude-Lorentz expression \cite{Jackson} of the permittivity, 
and also its quantum mechanical justification based on the electric dipole 
approximation \cite{CDL2}. In particular, the continuous superposition of 
resonances in (\ref{KKR}) describes a regime with absorption, while the 
quantum mechanics model is reduced to a discrete superposition 
of resonances and thus to the description of systems without absorption. 

In this section, it is proposed to express the new version of Kramers-Kronig 
relations for the electromagnetic Green's function or, equivalently, for the 
inverse operator $\He_e(z)^{-1}$. The objective is to 
transpose all the properties of the permittivity, and to make it possible to 
use all the knowledge on permittivity $\ep(\x,z)$ for the electromagnetic 
Green's function. 

The inverse operators $\He_e(z)^{-1}$ and $\He_0(z)^{-1}$ are expected to have 
properties similar to those of permittivities $\ep(\x,z)$ and $\ep_0$. Thus the 
following operator is considered:
\begin{equation}
\mathsf{R}(z) = \He_e(z)^{-1} - \He_0(z)^{-1} \, ,
\label{Rz}
\end{equation}
First, it is noticed that, as well as $\He_e(z)^{-1}$ and $\He_0(z)^{-1}$, 
the adjoint operator of $\mathsf{R}(z)^{-1}$ is 
\begin{equation}
\big[ \mathsf{R}(z)^{-1} \big]^\dagger = \mathsf{R}(- \overline{z})^{-1} \, ,
\label{Rdag}
\end{equation}
which is related to $\overline{\ep(z)} =\ep(- \overline{z})$. 
Next, let the operator $\mathsf{X}(t)$ be defined by 
\begin{equation}
\mathsf{X}(\t) = \dint_{\Gamma_\eta} dz \, \exp[-iz\t] \, {\mathsf{R}}(z) \, ,
\label{Chit}
\end{equation}
where $\Gamma_\eta$ is the horizontal line parallel to the real axis at a 
distance $\eta$, of complex numbers $z = \om + i \eta$ with $\eta>0$. 
It is stressed that this integral is well defined since, thanks to (\ref{Hzinfty}), 
${\mathsf{R}}(z)$ is bounded and decreases like $1/\om^2$. Also, this decrease 
in $1/\om^2$ implies that $\mathsf{X}(\t)$ is the Fourier transform of 
an integrable function, and thus $\mathsf{X}(\t)$ is continuous of $\t$.
The integral expression of $\mathsf{X}(\t)$ is independent of $\eta$ 
thanks to the analytic nature of the operator under the integral. 
The operator $\mathsf{X}(\t)$ is selfadjoint since, for $z=\om+i\eta$, 
\begin{equation}
\begin{array}{ll}
\mathsf{X}(\t)^{\dagger} &\hspace*{-1mm}= \dint_{\R} d\om \, \exp[i\overline{z}\t] 
\, {\mathsf{R}}(-\overline{z}) = \dint_{\R} d\om \, \exp[-iz\t] 
\, {\mathsf{R}}(z) \, , 
\end{array}
\end{equation}
where (\ref{Rdag}) has been used and the change $\om \rightarrow - \om$ has 
been performed to obtain the last expression. 
In addition, it can be checked that $\mathsf{X}(\t)$ vanishes for negative 
times. Indeed, in that case, the integral (\ref{Chit}) can be computed 
by closing the line $\Gamma_\eta$ by a semi circle with infinite radius in the 
upper half plane. Since all the functions are analytic, it is found that 
$\mathsf{X}(\t)=0$ if $\t<0$.
Hence, the quantity $\mathsf{X}(\t)$ is similar to the real susceptibility 
$\chi(\x,\t)$: it is selfadjoint and is associated with causality 
principle. 


Next, it is always possible to write for Im$z = \eta>0$
\begin{equation}
\mathsf{R}(z) = \dfrac{1}{2\pi} \dint_0^\infty d\t \, \exp[iz\t] \mathsf{X}(\t) \, .
\end{equation}
Using integration by parts, this expression above becomes
\begin{equation}
\mathsf{R}(z) = - \dfrac{1}{2\pi} \dint_0^\infty d\t \, \dfrac{\exp[iz\t]}{iz} 
\dt \mathsf{X}(\t)
\label{Rz2}
\end{equation}
since the values at the bounds vanish [$\mathsf{X}(\t)$ is continuous 
and vanishes at $\t = 0$]. Let $\xi = \nu + i \zeta$ be a complex number 
with positive imaginary part such that Im$(z) = \eta > \zeta > 0$. 
Then, equation (\ref{Rz2}) implies
\begin{equation}
\begin{array}{ll}
i\xi {\mathsf{R}}(\xi)
& = - \dfrac{1}{2\pi} \dint_0^\infty d\t \, \exp[i\xi\t] \dt \mathsf{X}(\t) 
\end{array}
\end{equation}
and the selfadjoint part is 
\begin{equation}
\begin{array}{l}
i\xi {\mathsf{R}}(\xi) - i \overline{\xi} 
{\mathsf{R}}(\xi)^\dagger
= - \dfrac{1}{2\pi} \dint_0^\infty \!\! d\t \, 
\exp[-\zeta\t] \, 2 \cos[\nu \t] \dt \mathsf{X}(\t) \\[4mm]
= - \dfrac{1}{2\pi} \dint_{\mathbb{R}} \!\! d\t \, \exp[i\nu\t]
\exp[-\zeta |\t|] \big\{ \dt \mathsf{X}(\t) 
+ \dt \mathsf{X}(-\t) \big\} \, . \\[4mm]
\end{array}
\end{equation}
The inverse Fourier transform is used: for $\t > 0$, 
\begin{equation}
\begin{array}{l}
\dfrac{1}{2\pi} 
\dt \mathsf{X}(\t) 
= \dint_{\mathbb{R}} \!\! d\nu \, \exp[-i\xi\t] \, 
\dfrac{\xi {\mathsf{R}}(\xi) - \overline{\xi} 
{\mathsf{R}}(\xi)^\dagger}{2 i \pi} \, .
\end{array}
\label{dtChit}
\end{equation}
Injecting this expression in (\ref{Rz2}) yields 
\begin{equation}
{\mathsf{R}}(z)
= - \dint_0^\infty d\t \, \dfrac{\exp[iz\t]}{iz} 
\dint_{\mathbb{R}} \!\! d\nu \, \exp[-i\xi\t] \, 
\dfrac{\xi {\mathsf{R}}(\xi) - \overline{\xi} 
{\mathsf{R}}(\xi)^\dagger}{2 i \pi} \, .
\label{Rz3}
\end{equation}
Next, by analogy with the function $\sigma(\x,\nu)$ defined by (\ref{sigma}), 
the following quantity is considered:
\begin{equation}
\mathsf{D}(\xi) = \dfrac{1}{2 i \pi} \big[ \xi {\mathsf{R}}(\xi) 
- \overline{\xi}{\mathsf{R}}(-\overline{\xi}) \big] \, ,
\label{Dxi}
\end{equation}
which is selfadjoint (and, contrary to $\sigma(\x,\nu)$, it is not positive 
since it is the difference of the one in the media minus the one in vacuum). 
The integral over the time in (\ref{Rz3}) is performed:
\begin{equation}
\begin{array}{ll}
{\mathsf{R}}(z)& = - 
\dint_{\mathbb{R}} d\nu \, \dfrac{1}{iz} \dfrac{-1}{i(z -\xi)} \, 
\mathsf{D}(\xi) \\[4mm]
& = - \dint_{\mathbb{R}} d\nu \, \dfrac{\mathsf{D}(\xi)}{z^2 -\xi^2} 
\, \dfrac{z + \xi}{z} \, .
\end{array}
\label{Xz3}
\end{equation}
The integral over $\nu$ in (\ref{dtChit}) and above is independent of the 
imaginary part $\zeta$ of $\xi$. Hence the integral in (\ref{dtChit}) and 
above can be considered in the limit $\zeta \downarrow 0$. Let $\mathsf{D}(\nu)$
be defined by 
\begin{equation}
\mathsf{D}(\nu) = \lim_{\zeta \downarrow 0} \mathsf{D}(\nu + i\zeta) \, ,
\end{equation}
it is left invariant when $\nu$ is changed in $-\nu$ 
[as well as in $\mathsf{D}(\xi)$]. It implies the following expressions
\begin{equation}
\begin{array}{rll}
{\mathsf{R}}(z)& = 
- \dint_{\mathbb{R}} d\nu \, \dfrac{\mathsf{D}(\nu)}{z^2 -\nu^2} 
\quad \quad & \text{Im}z>0 \, , \\[4mm]
z {\mathsf{R}}(z)& = 
- \dint_{\mathbb{R}} d\nu \, \dfrac{\mathsf{D}(\xi)}{z -\xi} 
& \text{Im}z> \text{Im}\xi \ge 0 \, .
\end{array}
\label{XKK}
\end{equation}
These equations define Kramers-Kronig expressions for the inverse 
Helmholtz operator. 
The first line of the equation above implies for the coefficients
\begin{equation}
\big\langle \phi, \He_e(z)^{-1} \psi \big\rangle = \big\langle \phi, \He_0(z)^{-1} \psi \big\rangle 
- \dint_{\mathbb{R}} d\nu \, \dfrac{\big\langle \phi, \mathsf{D}(\nu) \psi \big\rangle }{z^2 -\nu^2} \, , 
\label{HeKK}
\end{equation}
or, for the Green's function 
\begin{equation}
\G_e(\x,\y;z) = \G_0(\x,\y;z)
- \dint_{\mathbb{R}} d\nu \, \dfrac{\rho(\x,\y;\nu)}{z^2 -\nu^2} \, , 
\label{GeKK}
\end{equation}
where
\begin{equation}
\rho(\x,\y;\nu) = \dfrac{\nu \text{Im}[\G_e(\x,\y;z) - \G_0(\x,\y;z)]}{\pi} \,  
\label{rho}
\end{equation}
is related to the imaginary part of the relative Green's function [modulo the free 
Green's function $\G_0(\x,\y;z)$]. It is stressed that the quantity $\rho(\x,\y;\nu)$ 
is closely related to the local density of states which is just proportional to the 
trace of $\rho(\x,\x;\nu)$ [modulo the density of states in vacuum]. Hence the 
Kramers-Kronig expression (\ref{GeKK}) may be a first step for a generalization 
of the usual eigenmodes expansion of the Green's function \cite{RemiC15}. 
Indeed, consider the case of a closed cavity filled with a non dispersive and non 
absorptive dielectric media. Then, the Helmholtz equation without source term can 
be written 
\begin{equation}
\mathsf{L} \LE = z^2 \LE \, , \quad \mathsf{H}_e(z) = z^2 - \mathsf{L}
\label{ndna}
\end{equation}
where $\mathsf{L}$ is a selfadjoint operator independent of the complex frequency $z$. 
In the considered closed cavity, the operator $\mathsf{L}$ has a discrete set of 
eigenfunctions $| \phi_n \rangle $ and (real) eigenvalues $\om_n^2$
\begin{equation}
\mathsf{L} | \phi_n \rangle = \om_n^2 | \phi_n \rangle  \, ,
\label{ndna}
\end{equation}
which can be used to develop the inverse Helmholtz operator and the 
Green's function. Let the operator (\ref{Dxi}) act on
the eigenfunctions (here, for the sake of simplicty, the free inverse 
operator is not substracted):
\begin{equation}
\mathsf{D}(\xi) | \phi_n \rangle = \dfrac{1}{2 i \pi} \left[ \dfrac{\xi}{\xi^2 - \om_n^2}
- \dfrac{\overline{\xi}}{\overline{\xi}^2 - \om_n^2} \right] | \phi_n \rangle 
= \rho_n(\xi) | \phi_n \rangle \, ,
\label{Dxiphin}
\end{equation}
which, for $\xi = \nu + i\zeta$, leads to 
\begin{equation}
\rho_n(\xi) = \dfrac{1}{2 \pi} \left[ \dfrac{\zeta}{(\nu-\om_n)^2 + \zeta^2}
+ \dfrac{\zeta}{(\nu+\om_n)^2 + \zeta^2} \right] \, .
\label{Dxiphin2}
\end{equation}
When the imaginary part of $\xi$ tends to zero, the eigenvalue $\rho_n(\xi)$
becomes 
\begin{equation}
\lim_{\zeta \downarrow 0} \rho_n(\xi) = 
\rho_ n (\nu) = \dfrac{1}{2} \big[\delta(\nu - \om_n) + \delta(\nu + \om_n) \big] \, , 
\label{rho-n}
\end{equation}
and the operator $\mathsf{D}(\nu)$ can be written
\begin{equation}
\mathsf{D}(\nu) = \displaystyle\sum_n \rho_ n (\nu)  | \phi_n \rangle  \langle \phi_n | \, .
\label{Dnu}
\end{equation}
The function under the integral in (\ref{GeKK}) is then
\begin{equation}
\rho(\x,\y;\nu) = \displaystyle\sum_n \rho_ n (\nu)  \phi_n (\x) 
\otimes \overline{\phi_n(\y)} \, ,
\label{rho-nu}
\end{equation}
where the symbol $\otimes$ means that the tensor product is considered. Finally,
This expression is replaced into the equation (\ref{GeKK}) without the free part 
$\G_0(\x,\y;z)$:
\begin{equation}
\G_e(\x,\y;z) = - \displaystyle\sum_n \dfrac{1}{z^2 -\om_n^2} \, 
\phi_n (\x) \otimes \overline{\phi_n(\y)} \, , 
\label{GeKK-cav}
\end{equation}
Hence the classical expansion \cite{RemiC15} of the Green's function is 
retrieved. This confirms that the Kramers-Kronig expression (\ref{GeKK}) 
may be considered as an extension of the usual eigenmodes expansion 
\cite{RemiC15}. This Kramers-Kronig expression (\ref{GeKK}) may pave the 
way to obtain an eigenmodes expansion in the general case of 
frequency dispersive and absorptive systems. 

\section{Analytic properties in non dispersive systems\label{sec5}}

In non dispersive systems, the dielectric permittivity is independent 
of the frequency. Here, it is assumed that the dielectric constant is 
given by the expression (\ref{KKR}) of the permittivity where the frequency 
has been fixed to the real value $z = \om_0$: 
\begin{equation}
\ep(\x,\om_0) = \ep_0 - \dint_{\R} d\nu \, \dfrac{\sigma(\x,\nu)}{\om_0^2 - \nu^2} \, .
\label{ep-nd}
\end{equation}
The Helmholtz operator for the corresponding non dispersive system is 
\begin{equation}
\big[ \He_d(z) \LE \big] (\x) = z^2 \ep(\x,\om_0) \mu_0 \LE(\x) - \rot \rot \LE(\x) \, .
\label{H-nd}
\end{equation}

In this paper, it is proposed to exploit the new degree of freedom offered by the 
second complex frequency $\xi$ in $\He(z,\xi)$ to analyze the analytic 
properties of the non dispersive Helmholtz operator $\He_d(z)$. The 
identification of $\He_d(z)$ with the expression (\ref{Rezxi}) 
yields
\begin{equation}
z [ \ep(\x,\om_0) - \ep_0 ] = \xi [ \ep(\x,\xi) - \ep_0 ] \, .
\label{ep-iden}
\end{equation}
The permittivities are replaced by their Kramers-Kronig expressions (\ref{ep-nd}) 
and (\ref{KKR2}), which provides the following condition:
\begin{equation}
z \dint_{\R} d\nu \, \dfrac{\sigma(\x,\nu)}{\om_0^2 - \nu^2} \, = 
\dint_{\R} d\nu \, \dfrac{\sigma(\x,\nu)}{\xi - \nu} \, .
\label{ep-iden2}
\end{equation}
A sufficient condition to ensure this equation is 
\begin{equation}
z (\xi - \nu) = \om_0^2 - \nu^2 \Longleftrightarrow 
\xi = \nu + \dfrac{\om_0^2 - \nu^2}{z} \, .
\label{ep-iden2}
\end{equation}
Finally, the imaginary part of $\xi$ can be related to the one of $z$: 
\begin{equation}
\text{Im}(\xi) = \text{Im}(z) \, \dfrac{\nu^2 - \om_0^2}{|z|^2} \, .
\label{Im-xiz}
\end{equation}
In order to preserve the analytic properties of the inverse Helmholtz 
operator, it is necessary to have Im$(\xi)>0$. Since Im$(z)>0$, it is 
found that the difference $\nu^2 - \om_0^2$ has to be positive. This 
condition is realized if the function $\sigma(\x,\nu)$ vanishes for 
frequency $\nu$ smaller than a frequency $\nu_0 > \om_0$.
The resulting diectric constant is given by
\begin{equation}
\ep(\x,\om_0) = \ep_0 + \dint_{|\nu|\ge \nu_0} d\nu \, 
\dfrac{\sigma(\x,\nu)}{\nu^2 - \om_0^2} \, .
\label{ep-ndisp}
\end{equation}
The function under the integral is strictely positive and purely real. 
Thus the dielectric constant $\ep(\x,\om_0)$ is real, positive, and takes values 
greater than $\ep_0$: it describes a transparent dielectric. 

It has been shown that the analytic properties of the inverse Helmoltz operator 
can be preserved in non dispersive systems if the permittivity is the one of 
transparent dielectric. This analyticity property implies that such non dispersive 
systems are physically acceptable. Indeed, it can be checked that the causality 
principle is preserved for the solution $\E(\x,\t)$ of Maxwell's equation. 
Let $\J(\x,\t)$ be a current source switched on at $\t=0$, hence 
$\J(\x,\t) = 0$ for $\t<0$. Then, after the frequency decomposition, the 
current source
\begin{equation}
\LJ(\x,z) = \dint_0^{\infty} d\t \exp[izt] \, \J(\x,\t)
\end{equation}
is an analytic function in the upper half plane of the complex frequencies $z$. 
The time dependent electric field is given by
\begin{equation}
\E(\x,\t) = \dfrac{1}{2\pi} \dint_{\Gamma} dz \, 
\exp[-izt] \He_e(z)^{-1} (iz\mu_0) \, \LJ(\x,z) \, ,
\end{equation}
where $\Gamma$ is an horizontal line, parallel to the real axis, in the 
upper half plane. For negative times $\t$, the integral can be computed 
by closing the line $\Gamma$ by a semi circle with infinite radius in the 
upper half plane. Since all the functions are analytic, it is found that the 
electric field vanishes for negative times: $\E(\x,\t)=0$ if $\t<0$.
Hence, the causality principle is preserved. 

In addition, it can be checked that the light velocity $\v$ is always smaller 
than the one vacuum $\c$ since the dielectric constant takes values
larger than $\ep_0$:
\begin{equation} 
\v = \dfrac{1}{\sqrt{\ep(\x,\om_0) \mu_0}} \le \c = \dfrac{1}{\sqrt{\ep_0 \mu_0}} \, .
\label{vlec}
\end{equation}

Notice that the expression (\ref{ep-ndisp}) obtained for the permittivity, 
and derived from equation (\ref{ep-iden}), is just a sufficient 
condition which might be too strong. At this stage, this condition means that 
a physically acceptable description of a metallic behavior ($\om_0^2 > \nu^2$) or 
absorptive materials must include frequency dispersion. 

\section{Analytic properties with respect to the wavevector\label{sec6}}

In this section, it is assumed that a wavevector $\k$ can be defined, which 
requires for the geometry of the system to be invariant under a group of 
translations. The starting point is the expression of Maxwell's equations 
(\ref{Max-def}) introduced in section \ref{sec2C}. In the case the group 
of translations is discrete (periodic structure), the curl operator becomes 
$\rot + i \k \times$ after a Floquet-Bloch decomposition and, 
in case the group of translation is continuous (homogeneous structure), 
the curl operator becomes $i \k \times$ after a Fourier transform. The 
resulting free operator introduced in (\ref{Max-def}) becomes 
$\mathsf{M}_0(\k,z)$ while the potential $\mathsf{V}(\x,z)$ is left invariant. The 
wavevector $\k$ can have one, two or three components according to the 
geometry of the system which can be invariant under translations in one, 
two or three dimensions respectively. Let $\k'$ and $\k''$ be the real and 
imaginary parts of the wavevector: $\k = \k' + i \k''$. Then the ``imaginary'' part 
of the free operator $\mathsf{M}_0(\k,z)$ can be computed:
\begin{equation} 
\dfrac{\mathsf{M}_0(\k,z) - \mathsf{M}_0(\k,z)^\dagger}{2 i} = 
\left[ \begin{array}{lr}
\text{Im}(z) & - \ep_0^{-1} \k'' \times \\[1mm] 
\mu_0^{-1}\k'' \times & \text{Im}(z) \end{array} \right] \, .
\label{ImM0}
\end{equation}
The eigenvalues of this matrix are
\begin{equation} 
\lambda_0 = \text{Im}(z) \, , \quad 
\lambda_\pm = \text{Im}(z) \pm c \ks'' \, , 
\label{lambdaM0}
\end{equation}
where $\c$ is the light velocity in vacuum (\ref{vlec}), and 
$(\ks'')^2 = \k'' \cdot \k''$. Consequently, this imaginary part of 
the free operator is strictly positive as soon as 
\begin{equation} 
\text{Im}(z) \, > \, c \, \big| \ks'' \big| \, > \, 0 \, .
\label{domainzk}
\end{equation}
Since the imaginary part of the potential $\mathsf{V}(\x,z)$ is also positive (\ref{ImV}), 
the inverse $[\mathsf{M}_0(\k,z) + \mathsf{V}(\x,z)]$ is well-defined and analytic 
in the domain $\text{Im}(z) - c \, | \ks'' |  > 0$ of complex frequencies $z$ and 
wavevectors $\k$. And it is straightforward to extend this property to the inverse 
$[\mathsf{M}_0(\k,z) + \mathsf{V}(\xi,z)]$ in the domain defined by i) 
$\text{Im}(z) - c \, | \ks'' |  > 0$ and ii) $\text{Im}(\xi) > 0$. 

For instance, similar property has been used in \cite{Gra15} to derive new Kramers-Kronig 
relations for the reflection and transmission coefficients (via the Green's function) 
in the case of multilayered stacks illuminated with incident angle $\theta \neq 0$.
Indeed, in this situation, the square of the wavector appears to be 
$\k \cdot \k = (z^2 / \c^2) \sin^2 \theta$, which always meets the requirement 
(\ref{domainzk}).

\section{Conclusion\label{sec7}}

Analytic properties of the inverse Helmholtz operator and the Electromagnetic Green's 
function have been established. These properties are strongly related to the causality 
principle and the passivity requirement in frequency dispersive and absorptive media. 
Notably, the consequences for the permittivity make it possible to extend all the 
analytic properties of the free inverse Helmholtz operator (and the free electromagnetic 
Green's function) to the general inverse Helmholtz operator (and the Green's function). 
Hence the general inverse Helmholtz operator has been shown to be analytic in 
the domain $\text{Im}(z) - c \, | \ks'' |  > 0$ of complex frequencies $z$ and complex 
wavevectors $\k = \k' + i \k''$ (sections \ref{sec2} and \ref{sec6}). Moreover, it has 
been shown that an additional degree of freedom (the second frequency $\xi$) can be introduced 
in the inverse Helmholtz operator which remains analytic as soon as Im$(\xi) > 0$ 
(section \ref{sec2}). This additional frequency has been then exploited to retrieve 
that causal systems with non dispersive permittivity must have purely real dielectric 
constant taking values above the vacuum permittivity $\ep_0$ (section \ref{sec5}). 
Finally, asymptotic estimates and Kramers-Kronig expressions have been established for 
the inverse Helmholtz operator and the electromagnetic Green's function (sections \ref{sec3} 
and \ref{sec4}). Such Kramers-Kronig expressions can be considered as an extension of the 
well-known eigenmodes expansion of the Green's function in the case of a closed cavity 
filled with non dispersive and non absorptive media \cite{RemiC15}. 

It is stressed that all this results can be extended to magneto-electric materials 
\cite{GT10}. Indeed, in that case, it is enough to add the permeability $\mu(\x,z)$ in the 
expression of the matrix $\mathsf{V}(\x,z)$
\begin{equation}
\mathsf{V}(\x,z) = \left[ \begin{array}{lr}
z [\ep(\x,z) - \ep_0 ] & 0 \\ 0 & z [\mu(\x,z) - \mu_0 ] \end{array} \right] \, ,
\end{equation}
and then to apply the arguments proposed in this paper. Also, the analytic 
properties with respect to the wavevector can be used to extend the Kramers-Kronig 
expressions (like in section \ref{sec4}) to spatial dispersion. 

The results established in this paper may be used to calculate time-dependent 
electromagnetic fields \cite{GM12} and to establish rigorous eigenmodes
expansion in dispersive and absorptive systems. 


%
%

%

\begin{acknowledgments}
Anne-Laure Fehrembach and Gabriel Soriano (Institut Fresnel, Marseille, France) are 
acknowledged for fruitful discussions. 
\end{acknowledgments}

\section*{References}

%

\section*{Appendix}
\appendix
\section{Asymptotic behavior of the Green's function\label{A1}}
A proof of the asymptotic behavior (\ref{limzinfty}) is provided. 
As a first step, this asymptotic behavior is established for the 
inverse of the free Helmholtz operator given by
\begin{equation}
\big[ \He_0(z) \LE \big] (\x) = z^2 \ep_0 \mu_0 \LE(\x) - \rot \rot \LE(\x) \, . 
\label{He0}
\end{equation}
The Fourier transform with respect to the space variable, defined by 
\begin{equation}
\LE (\x) \longmapsto \widehat{\LE}(\k) = \dint_{\R^3} d\x \, 
\exp[- i \k \cdot \x] \, \LE(\x) \, , 
\label{Fouk}
\end{equation}
is performed to the free Helmholtz equation. The expression of $\He_0(z)$ 
in the Fourier space is 
\begin{equation}
\widehat{\He}_0(z) = 
z^2 \ep_0 \mu_0 - \ks^2 \big[ 1 -  \k \k / \ks^2 \big] \, ,
\label{He0k}
\end{equation}
where $\k\k$ is the tensor product of $\k$ with $\k$, and $\ks^2 = \k \cdot \k$. 
The inverse of the free Helmoltz operator is then
\begin{equation}
\widehat{\He}_0(z)^{-1} = \dfrac{1}{z^2 \ep_0 \mu_0} \, 
\big[ \k \k / \ks^2 \big]
+ \dfrac{1}{z^2 \ep_0 \mu_0 - \ks^2} \big[ 1 -  \k \k / \ks^2 \big] 
\label{Re0k1}
\end{equation}
or, equivalently,
\begin{equation}
z^2 \ep_0 \mu_0 \: \widehat{\He}_0(z)^{-1} = 1
+ \dfrac{\ks^2}{z^2 \ep_0 \mu_0 - \ks^2} 
\big[ 1 -  \k \k / \ks^2 \big] \, .
\label{Re0k2}
\end{equation}
Let $\phi$ and $\psi$ be square integrable fields, then the following 
coefficients are considered:
\begin{equation}
\big\langle \phi, \He_0(z)^{-1} \psi \big\rangle = 
\dint_{\R^3} d\k \, \overline{\hat\phi(\k)} \, 
\widehat{\He}_0(z)^{-1} \, \hat\psi(\k) \, .
\label{coeff0k1}
\end{equation}
The expression (\ref{Re0k2}) of the free resolvent leads to
\begin{equation}
z^2 \ep_0 \mu_0 \: \big\langle \phi, \He_0(z)^{-1} \psi \big\rangle  = 
 \big\langle \phi, \psi \big\rangle 
+  I(z) \, ,
\label{coeff0k2}
\end{equation}
with
\begin{equation}
I(z) = \dint_{\R^3} d\k \, \overline{\hat\phi(\k)} \, \dfrac{\ks^2}{z^2 \ep_0 \mu_0 - \ks^2} 
\big[ 1 -  \k \k / \ks^2 \big] \hat\psi(\k)  \, .
\end{equation}
In order to obtain the property (\ref{limzinfty}) for the free resolvent, 
it is enough to show that the number $I(z)$ tends to zero for large $|z|$. 
It is used that $|1 -  \k \k / \ks^2|$ is bounded by unity, 
and then the integral in 
\begin{equation}
\begin{array}{ll}
\big| I(z) \big| 
&\hspace*{-1mm} \le \dint_{\R^3} d\k \, \big| \hat\phi(\k) \hat\psi(\k) \big| 
\left| \dfrac{\ks^2}{z^2 \ep_0 \mu_0 - \ks^2} \right| \, ,
\end{array}
\label{Iz}
\end{equation}
is splitted in two parts: $I_+(z)$ for $|k|\ge K$ and $I_-(z)$ for $|k|\le K$. 
Writting the complex number as $z = |z| \exp[i\theta]$, it is obtained that 
$| z^2 \ep_0\mu_0 - \ks^2| \ge \ks^2 \sin^2 \theta$. The first part of the 
integral becomes
\begin{equation}
\big| I_+(z) \big| 
\le \dint_{|k|\ge K} d\k \, \big| \hat\phi(\k) \hat\psi(\k) \big| 
\dfrac{1}{\sin^2 \theta} \, ,
\label{Ipz}
\end{equation}
and, since the function $\hat\phi(\k) \hat\psi(\k)$ is integrable [i.e. in $L^1(\R^3)$], 
it can be made arbitrary small for large enough $K$. Next, the number $|z|$ is chosen large enough 
to have $|z|^2\ep_0\mu_0> K^2$, and the second part is bounded by 
\begin{equation}
\big| I_-(z) \big| \le \dfrac{K^2}{|z|^2\ep_0\mu_0 - K^2} 
\dint_{|k|\le K} d\k \, \big| \hat\phi(\k) \hat\psi(\k) \big| \, .
\label{Imz}
\end{equation}
This second part of the integral can be made arbitrary small when $|z| \rightarrow \infty$. 
Thus it is concluded that 
\begin{equation}
\lim_{|z| \rightarrow \infty} \, z^2 \ep_0 \mu_0 \, \big\langle \phi, 
\He_0(z)^{-1} \psi \big\rangle =  \big\langle \phi, \psi \big\rangle \, .
\label{H0zinfty}
\end{equation}

In order to extend this asymptotic behavior to $\He_e(z)^{-1}$, it is proposed 
in the second step to show that the coefficient 
\begin{equation}
C(z) = z^2\ep_0\mu_0 \, \big\langle \phi, \big[ \He_e(z)^{-1} - \He_0(z)^{-1} \big] \psi \big\rangle
\end{equation}
vanishes at the limit of infinite $|z|$. The second resolvent identity is used 
to express the difference
\begin{equation}
\He_e(z)^{-1} - \He_0(z)^{-1} = - \He_0(z)^{-1} z^2\mu_0 [\ep(\x,z) - \ep_0] 
\He_e(z)^{-1} \, .
\label{identity2}
\end{equation}
Let the function $\Psi$ be given by 
\begin{equation}
\Psi(z) = - z^2 \mu_0 [\ep(\x,z) - \ep_0] \He_e(z)^{-1} \psi \, .
\label{Psiz}
\end{equation}
It is well-defined for all $z$ since $z \He_e(z)^{-1}$
is bounded by $[\, \ep_0 \mu_0 \, \text{Im}(z) \,]^{-1}$ and, from (\ref{KKR2}), 
$z [\ep(\x,z) - \ep_0]$ is bounded by $[\dt \chi](\x,0^+) / \text{Im}(z)$. 
Also, the norm of $\Psi(z)$ can be made arbitrary small for 
large $|z|$ thanks to the estimate (\ref{z2epinfty}) for $z^2[\ep(\x,z) - \ep_0]$: 
\begin{equation}
\begin{array}{lcl}
\Arrowvert \Psi(z) \Arrowvert & \underset{|z| \rightarrow \infty}{\approx} &
[\dt \chi](\x,0^+) \,  \Arrowvert \mu_0 \He_e(z)^{-1} \psi \Arrowvert \\[2mm]
&\leq & \dfrac{[\dt \chi](\x,0^+)}{|z| \ep_0 \, \text{Im}(z)} \,  \Arrowvert \psi \Arrowvert \, .
 \end{array}
\label{Psizinfty}
\end{equation}
Finally, the preceeding equations (\ref{H0zinfty}--\ref{Psizinfty}) imply 
\begin{equation}
\begin{array}{lcl}
\big| C(z) \big| & = & z^2 \ep_0 \mu_0 \, 
\big| \big\langle \phi, \He_0(z)^{-1} \Psi (z) \big\rangle \big| \\[2mm]
& \underset{|z| \rightarrow \infty}{\approx} & \big| \big\langle \phi, \Psi (z) \big\rangle \big| 
\leq \dfrac{[\dt \chi](\x,0^+)}{|z| \ep_0 \, \text{Im}(z)} \, 
\Arrowvert \phi \Arrowvert \, \Arrowvert \psi \Arrowvert \, \\[4mm]
& \underset{|z| \rightarrow \infty}{\longrightarrow} & 0 \, .
\end{array}
\label{diff1}
\end{equation}
Hence the asymptotic behavior (\ref{limzinfty}) is proved. 
\end{document}